\documentclass{aa}  
\usepackage{epsfig,times,graphicx,bm,amssymb}
\usepackage{txfonts}
\usepackage{natbib}
\usepackage{setspace}
\usepackage{color}
\usepackage{ulem}

\begin{document} 

   \title{Stability of toroidal magnetic fields in stellar interiors}

	\titlerunning{Toroidal field stability} 
	\authorrunning{Ib\'a\~nez-Mej\'{\i}a \& Braithwaite}
	
   \author{J.C. Ib\'a\~nez-Mej\'{\i}a \inst{\ref{inst1}}$^,$\inst{\ref{inst2}}$^,$\inst{\ref{inst3}}\thanks{jibanezmejia@amnh.org}
          \and J. Braithwaite\inst{\ref{inst1}}}

   \institute{Universit\"at Bonn, Argelander Institut f\"ur Astronomie, Auf dem H\"ugel 71, 53121 Bonn, Germany \label{inst1}
              \and Universit\"at Heidelberg, Zentrum f\"ur Astronomie, Institut f\"ur Theoretische Astrophysik, Albert- Ueberle-Str. 2, 69120 Heidelberg, Germany  \label{inst2}
			  			  \and American Museum of Natural History, Department of Astrophysics, 79th Street at Central Park West, New York, NY 10024, USA \label{inst3}\\
}

%
%

  \abstract{}
  {Magnetic fields play an important role during the formation and evolution of stars. 
  Of particular interest in stellar evolution is what effect they have on the transport angular momentum and mixing of chemical elements along the radial direction in radiative regions. 
  Current theories suggest a dynamo loop as the mechanism responsible for maintaining the magnetic field in the radiative zone. This loop consists of differential rotation on one side and magnetohydrodynamic instability - the so--called Tayler instability - on the other.
  However, how this might work quantitatively is still an unsettled question, largely because we do not yet understand all the properties of the instability in question. In this paper we explore some properties of the Tayler instability.}
{We present 3D MHD simulations of purely toroidal and mixed poloidal-toroidal magnetic field configurations to study the behavior of the Tayler instability. 
For the first time the simultaneous action of rotation and magnetic diffusion are taken into account and the effects of a poloidal field on the dynamic evolution of unstable toroidal magnetic fields is included.
}
{In the absence of diffusion, fast rotation (rotation rate, $\Omega_{\parallel}$, compared to Alfv\'en frequency, $\omega_{\rm{A},\phi}$) is able to suppress the instability when the rotation and magnetic axes are aligned and when the radial field strength gradient $p<1.5$ (where $p\equiv\partial\ln B/\partial\ln \varpi$ and $\varpi$ is the cylindrical radius coordinate).
When diffusion is included, this system turns unstable for diffusion dominated and marginally diffusive dominated regions. 
If the magnetic and rotation axes are perpendicular to each other, $\Omega_{\perp}$, the stabilizing effect induced by the Coriolis force is scale dependent and decreases with increasing wavenumber.
In toroidal fields with radial field gradients bigger than $p>1.5$, rapid rotation does not suppress the instability but instead introduces a damping factor $\omega_{\rm{A}}/2\Omega_{\parallel}$ to the  growth rate, in agreement with the analytic predictions. 
 For the mixed poloidal-toroidal fields we find an unstable axisymmetric mode, not predicted analytically, right at the stability threshold for the non-axisymmetric modes; it has been argued that an axisymmetric mode is necessary for the closure of the Tayler-Spruit dynamo loop. }
{}

   \maketitle

\section{Introduction} \label{sec:introduction}

Rotation affects the structure of stars, inducing fluid motions such as circulation, and can enhance the mixing of chemical elements.
In stably stratified radiative interiors, however, fluid motions are largely restricted to isobaric surfaces, and strong turbulent fluid motions parallel to the isobars smooth velocity gradients in shells resulting in a shellular-like rotation \citep{Zahn92}.
In order to smooth the radial velocity gradients, hydrodynamic processes appear not to be able to transport enough angular momentum between neighboring shells. Consequently, it is thought that some process involving magnetic fields must account for any angular momentum transport in the radial direction.

 Even a relatively weak magnetic field is sufficient to couple different parts of the star and maintain a state of nearly uniform rotation. 
For the interior of the Sun, for example, a field of less than 1 gauss would be able to transmit the torque exerted by the solar wind through the interior \citep{Mestel53}.
Measurements of the solar core's rotation rate \citep{Chaplin01} show a very uniform rotation. 
This uniform rotation may be due to a magnetic field, but this field's origin, configuration, and strength are not known. 
By analogy with the magnetic A stars, one might speculate that a `fossil' magnetic field could exist in the core of the Sun. 
Since no significant net field is seen at the surface (averaged over the solar cycle), the radial component of such a fossil would, however, have to be weak -- on the order of a gauss or less. 
A field weaker than this will quickly wrap up into a predominantly toroidal field, under the action of the remaining differential rotation in the core. 
Eventually an instability will set in, limiting the growth of this toroidal field.

Near the end of the stellar lifetime, progenitors of white dwarfs and supernovae go through stages where the core contracts and spins up while the envelope expands and spins down.
The degree of coupling between core and envelope by a magnetic field in this stage will affect the rotation rates of the stellar leftovers, pulsars, and white dwarfs.
It is then fair to ask whether the rotation rate of the remnant is determined by the initial rotation of the progenitor, or if a secondary process must be responsible \citep{Spruit&Phinney98, Spruit98}.  

\citet{Tayler73} showed that any purely toroidal field should be unstable in at least some part(s) of the star. 
He derived stability conditions that are local in the meridional plane but global in the azimuthal direction; indeed, it is the low azimuthal $m$ modes that  become unstable first. 
Physically, one can think of the axisymmetric mode $m=0$ in the following way. 
A purely toroidal field $B_{\phi}=B_{\phi} (r,\theta)$, which must be axisymmetric in order to satisfy $\bm{\nabla}\cdot \textbf{B} = 0$, can be thought of as a collection of discrete neighboring circular flux tubes. 
Since all stars have a very high plasma-$\beta$, magnetically-induced motions are restricted to be almost incompressible, $\bm{\nabla}\cdot \textbf{v} \approx 0$. 
Now, any flux tube whose volume is held constant can reduce its energy by becoming shorter and fatter, which in the case of a flux tube encircling the axis of symmetry means contracting towards the axis. 
Obviously there is more energy to be released if the field is stronger, so that if we have two flux tubes (of equal volume) lying at different distances from the symmetry axis, if the outer tube contains a greater magnetic flux than the inner tube, interchanging the two will release energy.  
In a stellar radiative zone, the stable stratification largely suppresses motion in the radial direction, $\textbf{v}\cdot \textbf{g} \approx 0$. 
 However, to interchange flux tubes some radial motion is necessary; it is impossible for flux tubes confined to the same spherical surface to slip past one another.
Given that the unstable $m=0$ mode is axisymmetric, rotation, and thus Coriolis force, has no net effect on the stability conditions or growth rate, as it simply causes the annulus to rotate around the axis of symmetry.
 The non-axisymmetric modes are also driven by the reduction of energy by letting regions with strong magnetic field, and therefore a strong inwards-pointing curvature force, move inwards, and can be thought of in similar terms to the fluting instability in sunspots.

The growth timescale of the instability is on the order of the time taken for an Alfv\'{e}n wave to travel around the star on a toroidal field line. 
This timescale is very short compared to the evolutionary timescale of the star, e.g. about 10 years in a main-sequence star with a field of $1$ kG. 
This form of instability is likely to be the first to set in as the field strength of the toroidal field is increased \citep{Spruit99} but it is still uncertain what happens next. 
According to a scenario developed by \citet{Spruit02} the instability could lead to a regeneration of the poloidal field which will eventually be transformed into a toroidal field as a result of the remaining differential rotation, setting up a hydromagnetic dynamo loop. 
The balance between wrapping-up by differential rotation on the one hand and the destruction of the toroidal field by Tayler instability on the other determines the strength and configuration of the field.
The rate at which angular momentum and chemical elements are transported through the stellar interior depends on the resulting field strength, configuration and the details of the instability.
 This process has been implemented as a sub-grid model in stellar evolution calculations by \citet{Woosley02, Heger03} and \citet{Maeder03} and has been numerically reproduced in, with a very ideal configuration, simulations by \citet{Braithwaite06a}. More recently the process has apparently been seen in simulations by \citet{Rudiger2014}. \footnote{There have been claims that this Tayler-Spruit mechanism cannot work because of the lack of an axisymmetric mode; it turns out, however, that this is both misleading and untrue because (a) there is indeed a strong axisymmetric mode, as explained below, and (b) there is no convincing argument that an axisymmetric mode is necessary.}

In the long-term, we would like to determine the type of magnetic field that is maintained by differential rotation in a stably-stratified star, under the action of magnetic instabilities, and to develop a quantitative theory for the transport of angular momentum and chemical elements by magnetic fields in stars. This will require extensive simulations which include differential rotation and have a spherical geometry (Boldt \& Braithwaite, in prep.) In the meantime we aim to improve our understanding of the Tayler instability. Progress on this front has been made with local simulations by \citet{Braithwaite06b}, and has also been investigated in global simulations by \citet{Lander&Jones2011}. More recently \citet{Bonanno&Urpin2013a,Bonanno&Urpin2013b, Bonanno&Urpin2013c} have looked at various properties of the Tayler instability, in particular the effect of rotation, and find that rotation is essentially unable to suppress the instability; 
 one of our aims in this paper is to confirm this result with a different method and to ascertain the growth rate of the instability in the presence of fast rotation.

Another good reason to study the Tayler instability is to gain a better understanding of the range of magnetostatic equilibria available in radiative and other non-convective stars and zones. Since the discovery by \citet{Babcock47} of a strong magnetic field in the A star 78 Vir, strong, large-scale fields are now known to exist in a subset of early-type stars as well as white dwarfs and neutron stars (see e.g. Braithwaite \& Spruit subm. for a review). 
 These fields, known in the literature as fossil fields, are the relics of some previous stage of evolution, and have found their way into a stable equilibrium, evolving only on the very long diffusive timescale. An arbitrary initial magnetic field is known to relax in a non-convective star into such an equilibrium \citep{Braithwaite&Spruit04,Braithwaite&Nordlund05}, and there are apparently a large range of equilibria available, including both approximately axisymmetric 'twisted torus' configurations as well as more complex non-axisymmetric configurations \citep{Braithwaite08}. Amongst the axisymmetric equilibria, it is an interesting question to ask in what ratio the poloidal and toroidal field components must be present to ensure stability, since both are unstable on their own. It seems that a relatively weak poloidal field can stabilize a toroidal field \citep{Braithwaite09, Akgun13} but the detail of how this works is still uncertain. One reason why the ratio of the two is particularly interesting is that a star with a predominantly poloidal field is oblate in shape, but if the field is predominantly toroidal, the star is prolate. Fast-spinning highly-magnetized neutron stars (the so-called millisecond magnetars) with predominantly toroidal magnetic fields might flip over until the magnetic and rotation axes are orthogonal, the damping of free precession reducing the kinetic energy by maximizing the moment of inertia about the rotation axis. Such stars would emit gravitational waves potentially detectable with the next generation of detectors.

In this paper we use numerical methods to investigate the linear properties of the Tayler instability, much of which is still uncertain. We provide a useful check on the analytical results, which may have missed some aspects of the mechanism. 
We use similar methods to those used by \citet{Braithwaite06b} to investigate the Tayler instability, but we investigate different aspects of the instability. 
In the next section, we look at what is known about the instability, before describing our numerical model in section \ref{sec:model}, presenting our results in section \ref{sec:results} and summarizing in section \ref{sec:conclusions}.

\section{Stability of toroidal fields: analytic results}  \label{sec:analytics}

\citet{Tayler73} derived the stability conditions of purely toroidal fields in radiative  stellar interiors, ignoring the effects of viscosity and of thermal and, magnetic diffusion. 
The occurrence of instability for the adiabatic case is independent of the field strength but only depends on the field configuration. 
However, when the instability is already present the growth rate depends on some measure of the field strength and the stratification.
The field strength is characterized by the Alfv\'en frequency, which can be thought as the inverse of the timescale taken by an Alfv\'en wave to go around the star along a toroidal field line, defined as
\begin{equation}
	\omega_{\rm{A}}=\frac{\textrm{v}_{\rm{A}}}{R_\ast}=\frac{\rm{B}_{\phi}}{R_\ast \sqrt{4\pi\rho}},
	\label{eq:tayler_inst}
\end{equation}
 where $\textrm{v}_{\rm{A}}= \rm{B}_\phi /\sqrt{4\pi\rho}$ is the Alfv\'en velocity in a toroidal field line and $R_\ast$ is the radius of the star. 
The stratification is characterized by the Brunt-V\"ais\"al\"a or the buoyancy frequency, $N$, and radiative zones are always in the regime
\begin{equation}
	N \gg \omega_{\rm{A}, \phi}.
\end{equation}
This is the same as saying that the thermal energy density is much bigger than the magnetic energy density, i.e.  a large plasma- $\beta=\rm{P}/8\pi \rm{B}^2$ parameter, which is true in all stars.

Following Tayler, we make the analysis easier by using cylindrical coordinates $(\varpi,\phi,z)$.
We now consider an axisymmetric, azimuthal toroidal field independent of height $z$, $\textbf{B}=\textrm{B}(\varpi)\bf{e}_\phi$, and look at the region around the axis.
The important parameter is
 the radial field gradient $p$:
\begin{equation}
	p\equiv\frac{\rm{d}\log \rm{B}}{{\rm d}\log \varpi}   \quad\quad\quad  \text{as}  \quad\quad\quad  \varpi\rightarrow 0.
	\label{eq:field_grad}
\end{equation}

\begin{figure}
	\centering
	\includegraphics[scale=0.58]{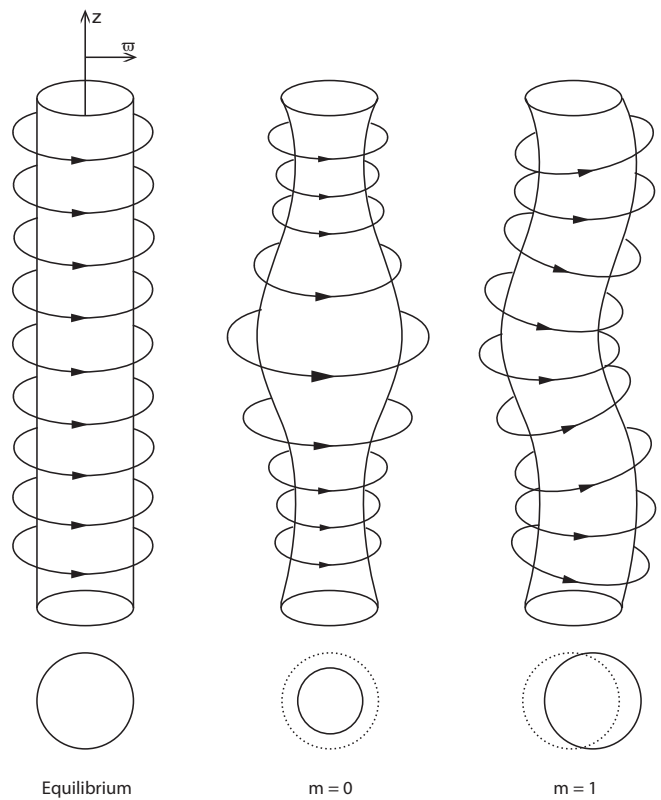} 
	\caption{Physical form of the different unstable azimuthal $m$ modes present in a plasma column with a purely toroidal field. Below each mode the cross section of the co-moving surface is shown along with the cross section of the equilibrium state. The only axisymmetric mode is $m=0$ and the only mode that preserves the column's cross section is the $m=1$ mode.}
	\label{fig:tayler_inst}
\end{figure}

Now, if we perturb  a toroidal field in equilibrium, the resulting displacement can be described by
\begin{equation}
	 \xi \sim \it{e}^{i(m\phi+kz)+\sigma t},
	 \label{eq:displacement}
\end{equation}
where the variation in the $\varpi$ direction is not included as its length scale is much larger.
Although displacements are predominantly horizontal because of the vertical stratification, the incompressible continuity equation (whose use is justified by the very high plasma- $\beta$) shows that a horizontal displacement $\xi_{\rm{h}}$ is necessarily accompanied by a radial displacement on the order of $\xi_r \sim \xi_{h} l_{r} / l_{h}$, where $l_h$ and $l_{r}$ are horizontal and radial length-scales, respectively, so that near the axis $l_{r}=1/k$.
The shape of some of the unstable modes is shown in figure \ref{fig:tayler_inst}. 
Note that the $m=0$ mode is also known as the sausage mode and the $m=1$ mode as the kink mode. 

In this figure we can observe why perturbations like (\ref{eq:displacement}) in a purely toroidal field are unstable. 
Regions where magnetic field is pinched have higher magnetic pressure forcing the fluid to displace opposite to the pressure gradient, leading to a runaway process.
Figure \ref{fig:tayler_inst} shows the equilibrium configuration and the shape of the azimuthal $m=0$ and $m=1$ unstable modes.
In a stellar interior we often expect a vertical stratification, and vertical displacements experience a damping effect because they have to do work against the stratification. 
However, given that the $m=1$ mode mostly drives azimuthal displacements, the damping effect of the stable stratification is very weak.

\citet{Tayler57} showed that for the instability to develop the field must satisfy
\begin{equation}
	 p>\frac{m^2}{2}-1 \,\, (m \neq 0) \,\,\, \text{and} \,\,\, p>1 \,\,\, (m=0).
	 \label{eq:stability_non-rotating}
\end{equation} 
These instability conditions are valid only for non-viscous, non-rotating, non-diffusive, radiative stellar interiors.
Examining this stability condition (\ref{eq:stability_non-rotating}) we observe that for a $p=1$ field, the $m=1$ mode is the only unstable mode, the $m=0$ and $2$ modes are marginally stable and the other modes are stable.
For a somewhat steeper radial field gradient of $p=2$, modes $m=0, 1, 2$ are expected to be unstable.

To study the growth of the instability we define the local toroidal\footnote{For a purely toroidal field the toroidal and total (that is, replacing $B_\phi$ by $B$ in the definition) Alfv\'en frequencies are the same; however for mixed poloidal-toroidal fields this distinction is necessary.}  Alfv\'en frequency as
\begin{equation}
	\omega_{\textrm{A}, \phi} \equiv \textrm{v}_{\textrm{A},\phi}/\varpi = \frac{B_{\phi}} {\varpi \sqrt{4\pi\rho}},
	\label{eq:alfven_freq_toroidal}
\end{equation}
where $\textrm{v}_{\textrm{A},\phi}$ is the toroidal Alfv\'en speed and $\varpi$ the distance to the axis, we expect the growth rate of the instability to be
\begin{equation}
	\sigma \sim \omega_{\textrm{A}, \phi}.
	\label{eq:ideal_growth}
\end{equation}
Given that toroidal magnetic fields in stars are buried under the surface, there is no direct observational evidence of their configuration, therefore we have no knowledge of the radial magnetic field gradient.
Even less understood is the configuration of a magnetic field inside a differentially rotating star.
However we know that towards the magnetic axis the field must tend to zero, and to avoid singularities the radial gradient  $p \geq 1$ for $\varpi \rightarrow 0$, and for this reason in this paper we start by studying the $p=1$ case. 
Since it is expected that there is a qualitative threshold at $p=3/2$ (see section \ref{subsec:rotation_effects}), we then also study the case of a $p=2$ field gradient. 

The unstable growth rate in equation \ref{eq:ideal_growth} is valid only for purely toroidal, non-rotating, non-stratified and non-diffusive cases. 
We now discuss the behavior of the Tayler instability for situations where these very ideal conditions are relaxed and discuss what happens when magnetic diffusion is included, rotation is present and the magnetic field configuration is not purely toroidal anymore.
By doing so, the stability conditions and growth rates are modified, and furthermore growth rates develop length scale dependencies.
A stable stratification has major effects on the behavior of the instability, this is however outside the scope of this paper and is already discussed in \citet{Braithwaite06b}.

\subsection{Effects of rotation} \label{subsec:rotation_effects}

All stars are rotating and a big fraction rotates at very high speeds \citep{Collins63, Ramirez-Agudelo}.

\subsubsection{Rotation and magnetic axes aligned}

\citet{Acheson78,Pitts&Tayler85} and \citet{Spruit99} showed that both the stability conditions and the unstable growth rate are affected by fast rotation. 
When the rotation frequency is higher than the Alfv\'en frequency ($\Omega_{\parallel} > \omega_{\rm{A}, \phi}$) and the rotation and magnetic axes are aligned, the stability condition is modified from (\ref{eq:stability_non-rotating}) to
\begin{equation}
 p>\frac{m^2}{2}+1 \;\; ( m \neq 0 ) \quad\quad\quad \text{ and } \quad\quad\quad p>1 \;\; (m=0) 
 \label{eq:stability_rotating}
\end{equation} 
so that the condition of stability against all $m$ modes is $p<3/2$ rather than $p<-1/2$ as before, and furthermore, the growth rate when any unstable mode is present is damped by a factor of $\omega_{\textrm{A}, \phi}/\Omega_{\parallel}$ \citep[see][]{Acheson78,Spruit99}. 

\begin{figure} 
	 \centering
		 \includegraphics[scale=0.5]{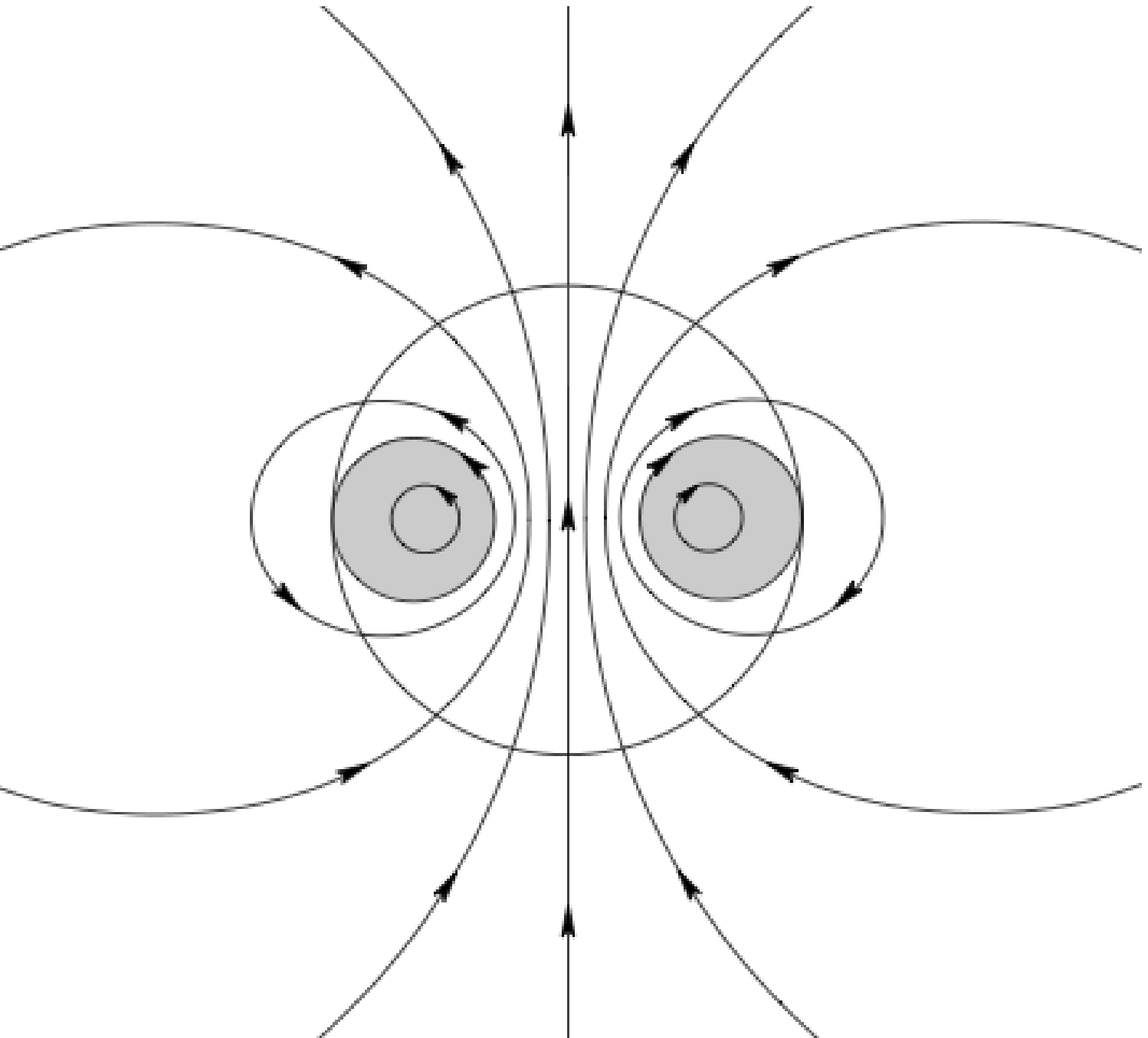}
		 \includegraphics[scale=0.35]{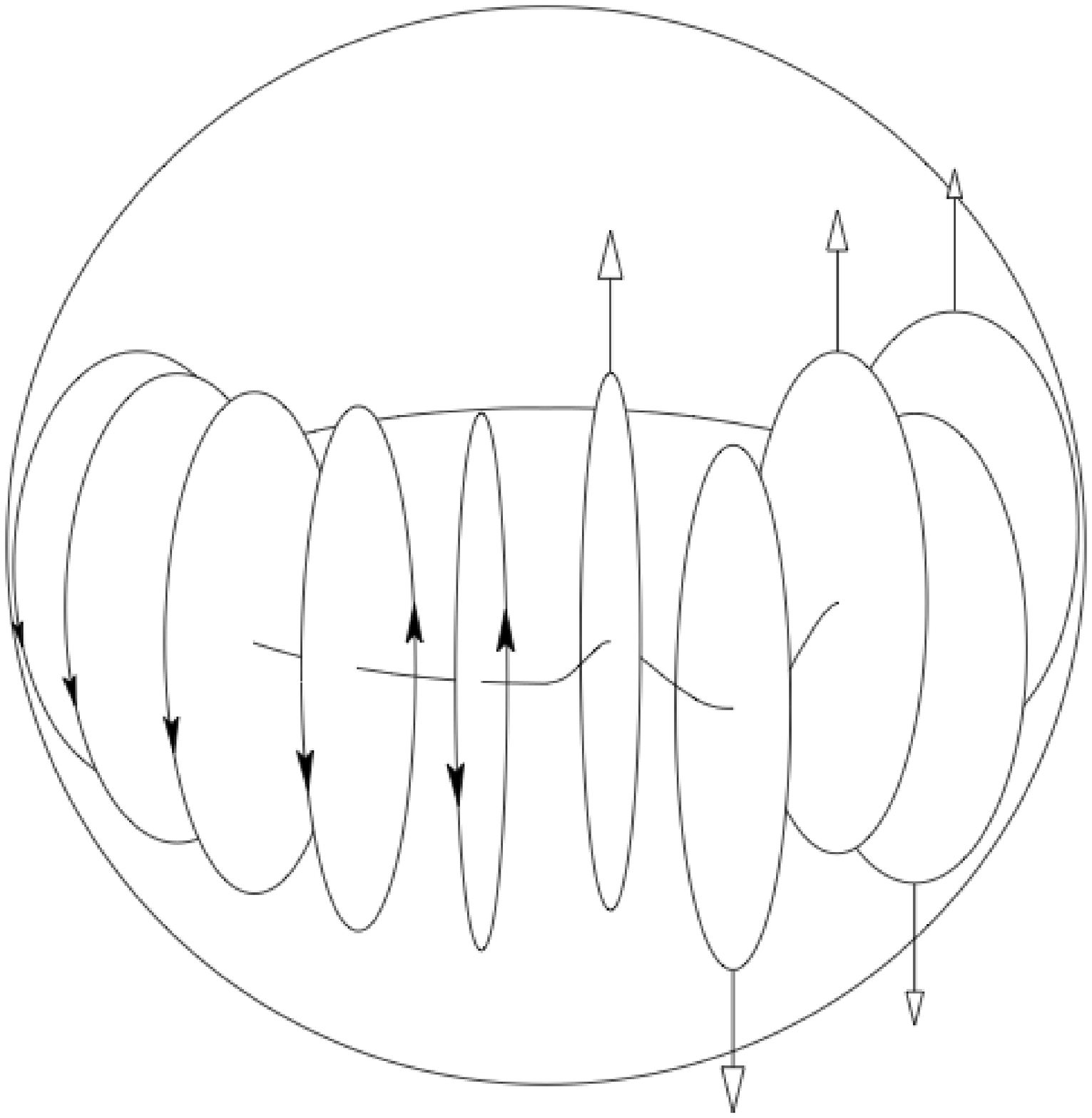}
	  \caption{{\it Above:} Mixed poloidal-toroidal field configuration. 
	  If the toroidal field is too weak or indeed absent, the poloidal field is subject to an instability. 
	   {\it Below:} Left, an equilibrium configuration with poloidal field loops threaded by the neutral line, and on the right-hand-side the kink-like instability present around the neutral line.
	  Unstable displacements are restricted to be approximately perpendicular to the stratification. Figure taken from \citealt{Braithwaite07}. }
	 \label{fig:poloidal_field}
\end{figure}

\subsubsection{Rotation and magnetic axes not aligned}

It is not always the case that the magnetic axis, the rotation axis and the gravity are aligned, as in the cases looked at above. Figure \ref{fig:poloidal_field} (upper panel) shows the poloidal field lines in an axisymmetric configuration, and a toroidal-field-like configuration is observed around the neutral field line around the star,  where the poloidal field vanishes.
This configuration suffers from the same kind of instability (Tayler instability when considered locally), the difference being that the magnetic axis (in this case the neutral line) is perpendicular to gravity, but because of the switch in geometry, it is now the poloidal field that drives the instability, instead of the toroidal field as before; during local considerations it therefore makes sense to switch the definitions of the two words. Also note that in the presence of rotation, the magnetic and rotation axis wont be alligned somewhere in the star.
 Kink-like instabilities appear with displacements perpendicular to the direction of stratification, $ \textbf{v} \cdot \textbf{g} \approx 0$, as appreciated in figure \ref{fig:poloidal_field} (lower panel). 
Displacements are mainly in the direction of the arrows, but, as discussed in section \ref{sec:analytics}, there are also necessarily smaller displacements in the direction parallel to the neutral line .
The ratio between the displacements in these two directions should be equal to the ratio of length scales in the same two directions, as we can see from the constraint $\bm{\nabla} \cdot \textbf{v} \approx 0$, a consequence of the high plasma-$\beta$. 

The Coriolis force affects motions perpendicular to the rotation axis; in the case where the magnetic and rotation axes are aligned, unstable displacements are mainly perpendicular to the rotation and magnetic axes, and stabilization in the case $-1/2<p<3/2$ is achieved when $\Omega_{\parallel} \geq \omega_{\textrm{A}, \phi}$. 
When the magnetic and rotation axes are not aligned, the amplitude of the displacements perpendicular to the rotation axis  decrease with increasing inclination angle, and the extreme case occurs when the rotation and magnetic axes are perpendicular to each other.
Then naively we think that in order to stabilize the perpendicular rotator, a faster rotation rate is required, and as we assumed that the fluid is almost incompressible, $\bm{\nabla} \cdot \textbf{v}\approx 0$, we derive the stability condition
\begin{equation}
	\Omega_\perp \geq \omega_{\textrm{A}, \phi}  \varpi_{0} k,
	\label{eq:critical_rotation_oblique}
\end{equation}
where $\Omega_{\perp}$ is the perpendicular rotation rate with respect to the magnetic field axis, $\varpi_{0}$ is the typical length scale of the motion perpendicular to the magnetic field and $k$ is the wavenumber in the direction of the magnetic axis (here the neutral line) as defined in (\ref{eq:displacement}).
Note that the new perpendicular rotating stability conditions now dependent on the wavenumber, unlike in the aligned rotator case.

\begin{figure}[!h]
	\centering
  \includegraphics[scale=1.00]{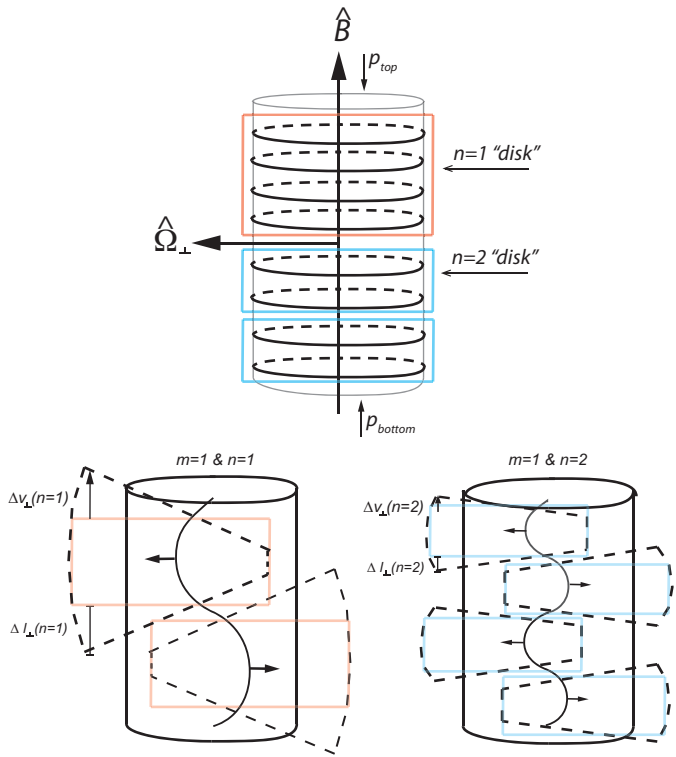}
 \caption{Representations of the first two vertical $n$ modes of the unstable $m=1$ mode.
 The figure on top shows the purely toroidal magnetic field configuration, where the orientation of the magnetic symmetry axis $\hat{B}$ is perpendicular to the rotation axis $\hat{\Omega}_{y}$. It also shows what an $n=1$ and $n=2$ disk looks like as the red and blue box, respectively.
 The two figures below show the slipping disks for the $n=1$ (left) and $n=2$ (right) cases. The magnitude of the deformation $\Delta l_{\perp}$, and the deformation velocity $\Delta \textrm{v}_{\perp}$, suffered by the disks is different depending on the unstable mode.
 Small $n$ modes present large deformations, whereas high $n$ modes show small deformations, $\Delta l_{\perp}(n=1) > \Delta l_{\perp} (n=2)$ and $\Delta \textrm{v}_{\perp}(n=1) > \Delta \textrm{v}_{\perp} (n=2)$.  }
 \label{fig:oblique_deformation}
\end{figure}

This effect can be explained if we imagine the plasma column to be a series of disks subject to pressure on top and below.
When the disks displace perpendicular to the magnetic axis they suffer a deformation fattening at one side and shrinking at the other.
The magnitude of the deformation depends on the size of the disks, as shown in figure \ref{fig:oblique_deformation}, where disk of different sizes correspond to different vertical $n$ modes.
 This is simply a consequence of the constraint ${\bm\nabla}\cdot \textbf{v}\approx0$, which can be written as $k \textrm{v}_z\sim \varpi_0 \textrm{v}_\varpi$.
The Coriolis force affects motion in the vertical direction; Greater deformations, in other words small $\textit{n}$ modes, feel a stronger stabilizing effect.
 Note that the Coriolis force also affects motions perpendicular to both the magnetic and rotation axis, but this motion, at least for the $m=1$ mode, can be minimized by combining left- and right-handed spiral modes with the right phase.

\subsection{Effects of magnetic diffusion} \label{subsec:mag_diff_effects}

Until now we have only considered adiabatic perturbations, that is ignoring the effects of thermal and magnetic diffusion, as well as viscosity. 
We now relax the flux freezing condition and allow the field lines to slip through the plasma. 

\citet{Pitts&Tayler85} and \citet{Spruit99} suggested that in the presence of any diffusive process, thermal or magnetic, the instability condition in fast rotators (\ref{eq:stability_rotating}) would recover its original form given in condition (\ref{eq:stability_non-rotating}) \citep[see][Appendix]{Spruit99}.
The reason for this can be thought of in terms of energy dissipation: rotation tends to give stability because the Coriolis force, which acts perpendicularly to the velocity and has consequently no effect on the energies, drives the fluid back into its equilibrium position; diffusion damps these epicycle motions, making a return to the equilibrium energetically impossible; as magnetic energy is turned into heat the system moves irreversibly away from its original equilibrium.

Magnetic diffusion is scale dependent, and therefore when included in the analysis, the unstable growth rate of the Tayler instability becomes scale dependent as well \citep[see e.g.][]{Spruit99}.
For a given magnetic diffusivity, $\eta$, the diffusion rate is given by, $\eta k^2$, where $k$ is the wavenumber.%
At first glance one expects this to impose an upper limit for the unstable wavenumbers that are able to grow; perturbations at a larger wavenumbers should be smoothed away faster than they can grow. 
We have stability if
\begin{eqnarray}
 \eta k^2 > \sigma
	\label{eq:max_wavenumber}.
\end{eqnarray}
Therefore for a given wavenumber we can define a critical diffusivity 
\begin{equation}
	\eta_{\rm crit} = \omega_{A,\phi} /k^2,
\end{equation}
From now on, we will express the different diffusive environments as fractions of the critical magnetic diffusivity
\begin{equation}
	n_{\eta} = \frac{\eta}{\eta_{crit}},
	\label{eq:magdiff_frac}
\end{equation}
where one can always move back and forth between fraction of critical magnetic diffusivity and the wavenumber of interest with the following relation
\begin{equation}
	n_{\eta} = k^2 \left(\frac{\eta}{\omega_{A,\phi}}\right),
	\label{eq:n_eta}
\end{equation}
where $\eta$ is the magnetic diffusivity of the system and $k = 1/l_{z}$ is the vertical wavenumber.

\subsection{Mixed poloidal-toroidal configurations} \label{subsec:mixed_field_effects}

The addition of a poloidal field is expected to modify or suppress the instability. If one imagines the toroidal and poloidal fields to operate independently of each other (in some vague sense) then some of the energy liberated from the toroidal field must be used to bend the poloidal field lines. If the poloidal field is strong, the instability might be suppressed altogether. However, the effect will depend on the wavenumber of the mode; whilst all wavenumbers release the same energy from the toroidal field, higher wavenumbers have to do more work against the poloidal field. This gives rise to a wavenumber stability threshold. 

In general, if an instability displaces a fluid element horizontally a distance $\xi_{\textit{h}}$, it does so by means of a force per unit mass of $\textit{F}_{\textit{h}}=\sigma^2\xi_{\textit{h}}$ where $\sigma$ is the growth rate. In the case of the Tayler instability, $\sigma\approx \omega_{\textrm{A}, \phi}$ as discussed at the beginning of section 2 (equation \ref{eq:tayler_inst}).
Imagine, in the cylindrical geometry used above, that there is a poloidal field in the $z$ direction. The displacement caused by the instability drags the poloidal field with it, which reacts against this bending with a force per unit mass of $B_z B_\varpi k/\sqrt{4\pi\rho}\approx B_z^2 \xi_{\rm h} k^2/\sqrt{4\pi\rho}$, where $k$  is the radial (vertical) wavenumber. 
For a perturbed displacement to grow it is required that the instability force is bigger than the restoring force by bending the poloidal field line,
 which gives a condition on the unstable wavenumbers
\begin{equation}
k < \frac{\sigma}{\textrm{v}_{{\rm A},z}},
	\label{eq:poloidal_stability}
\end{equation}
where $\textrm{v}_{{\rm A},z}=B_z/\sqrt{4\pi\rho}$ is the vertical Alfv\'en speed. 
See \citet{Spruit02} and \citet{Braithwaite09} for a fuller discussion of this topic.

\section{Model} \label{sec:model}

\begin{figure}
	\centering
	\includegraphics[scale=0.40]{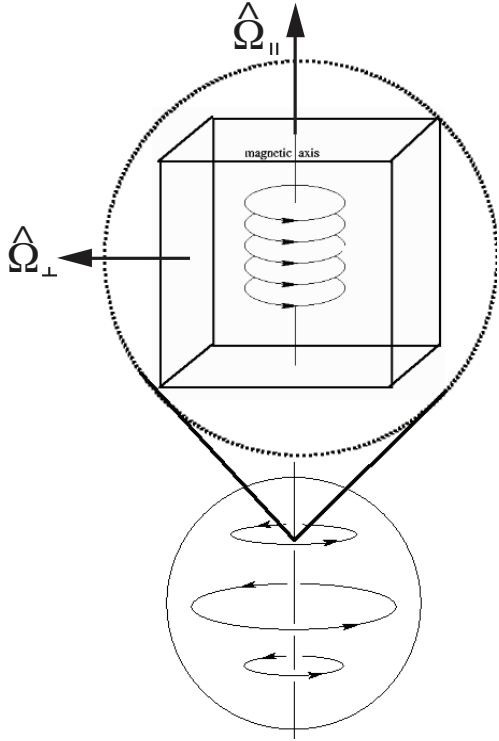} 
	\caption{Predominantly toroidal magnetic field configuration in the interior of a star. The box shows a zoomed-in region containing the toroidal field lines around the magnetic axis and bold arrows showing the direction of the parallel and perpendicular rotation axes.}
	\label{fig:Simulation_box}
\end{figure}

To determine the effects of rotation and magnetic diffusion on the Tayler instability, local simulations of a radiative stellar interior are performed. 
For our local simulations we select a box containing the magnetic axis of the star, located at the center of the box, as shown in figure \ref{fig:Simulation_box}. 
Global simulations of the stellar radiative region are not suitable at this stage because of resolution limitations to accurately resolve small scales. 
However, global simulations might eventually be necessary to investigate the non-linear behavior in a differentially rotating star, or rather, global in two dimensions and local in the radial direction.

The equation of state used is the ideal equation of state, in units such that the molar gas constant divided by the molecular mass is unity
\begin{eqnarray}
	\textit{P}=\rho\textit{T}	\;\;\; \text{and} \;\;\; \textit{e}=\frac{P}{\gamma-1}.
\end{eqnarray}

To trace the evolution of the system we solve the standard MHD equations, 
momentum equation:
\begin{eqnarray}
	\frac{\partial{\rho\textbf{v}}}{\partial t}=-{\bm\nabla}\cdot(\rho\textbf{vv})-{\bm\nabla} P+ \frac{1}{4\pi}{\bm\nabla}\times {\bf B} \times {\bf B} -\textbf{g}\rho+2\rho\textbf{v}\times\Omega, 
\label{eq:momentum_eq}
\end{eqnarray}
continuity equation,
\begin{eqnarray}
	\frac{\partial{\rho}}{\partial t}=-\nabla \cdot \rho \textbf{v}, 
	\label{eq:conservation_of_mass}
\end{eqnarray}
induction equation,
\begin{eqnarray}	
	\frac{\partial{\textbf{B}}}{\partial t} = \nabla\times \left[ (\textbf{v}\times\textbf{B})-\eta\nabla\times\textbf{B} \right], 
	\label{eq:induction2}
\end{eqnarray}
and the specific internal energy:
\begin{eqnarray}
	\frac{\partial e}{\partial t}=-\nabla\cdot(\textit{e}\textbf{v}) - P\nabla\cdot\textbf{v} 
	\label{eq:energy} %
\end{eqnarray}
where $\rho$ is the density, $\textbf{v}$ the velocity, $\textbf{B}$ the magnetic field, $\eta$ the magnetic diffusivity, $\textit{e}$ the internal energy, $\textbf{g}$ the gravity, $\textit{P}$ gas pressure, and $\textit{T}$ gas temperature. The ratio of specific heats  $\gamma$ is set to $5/3$. 
We note that in the momentum equation (\ref{eq:momentum_eq}) we neglected viscosity and the Coriolis term is explicitly included. 
In the induction equation (\ref{eq:induction2}) magnetic diffusion $\eta$ is spatially constant. 
The energy equation neglects viscous and Ohmic heating and thermal conduction, since we are not interested at this stage in viscous and thermal effects. 
We do not include the effects of gravity in this study.

\subsection{Numerical code} \label{subsec:numerical_code}

We use the stagger code, a parallel 3D MHD code originally developed by \citep{stagger}, and continuously improved over the years by its user community. 
The Stagger-code uses a sixth-order explicit finite-difference scheme for numerical derivatives and the corresponding fifth- order interpolation scheme. 
The solution of the hydrodynamic equations is advanced in time using an explicit third-order Runge-Kutta integration method \citep{Williamson80}.
The code operates in a staggered, Eulerian, Cartesian mesh, which seem to be a better implementation than cylindrical and spherical coordinates, as a Cartesian mesh avoids the coordinate singularity in the axis present in cylindrical and spherical grid-based codes.
Another advantage of a Cartesian based code is reflected in the simplicity of the equations with respect to another coordinate system.
Cartesian coordinates also have their own disadvantages such as the noise introduced by fitting a cylindrical shape, e.g. torus, in squared boxes, or the extra computational time cost due to the un-used corners of the box. \\

For the simulations of purely toroidal fields a resolution of $48^2$ in the azimuthal direction and 32 in the vertical direction was used, for the mixed poloidal-toroidal fields the resolution was $48^3$. Test cases with a resolution of $96^3$ were performed in order to compare the results with the low resolution simulations and no significant differences were observed. Periodic boundaries are used in all simulations; see below.

The code also contains highly localized diffusivities that prevent the growth of wavelengths close to the Nyquist frequency. 
We however decided to switch off this artificial diffusive terms in order to avoid non-physical effects. 
This was possible given the clean initial conditions, where the development and growth of the Tayler instability occurs at very early times when no artificial stabilization is required.

\subsection{Initial conditions} \label{subsec:initial_conditions}

The initial conditions correspond to the equilibrium configuration of a plasma containing a magnetic field of the form:
\begin{eqnarray}
 \textbf{B}=\textrm{B}_\phi(\varpi)\textbf{e}_\phi+\textrm{B}_z\textbf{e}_z.
  \label{eq:magnetic_field}
\end{eqnarray}
The magnetic configuration in this paper has no radial component, $\textrm{B}_{\varpi}=0$, and is a combination of an azimuthal, $\textrm{B}_{\phi}$, and a vertical, $\textrm{B}_z$, components with no dependence on height $z$ or angle $\phi$. 

The azimuthal component is an axisymmetric torus defined by
\begin{eqnarray} \label{eq:toroidal_field}
 	B_{\phi}(\varpi)=B_{0,\phi}\left(\frac{\varpi}{\varpi_0}\right)^p e^{-(\varpi/\varpi_0)^2} ,
 	\label{eq:bphi}
\end{eqnarray}
where $\varpi_{0}$ is a constant length, set to one quarter of the horizontal size of the box, $\textit{p}$ is the radial field gradient, set to either 1 or 2 in the simulations, and $\textrm{B}_{0,\phi}$ is a constant. The exponential term $\exp(\varpi^2$/$\varpi_0^2)$ shapes the field such that it behaves as $\textrm{B}\propto \varpi^{\textit{p}}$ near the axis and vanishes farther away so as to avoid problems at the boundaries.
The local toroidal Alfv\'en frequency, as defined in (\ref{eq:alfven_freq_toroidal}), is $\omega_{\textit{A},\phi} = \textrm{B}_\phi/\varpi\sqrt{4\pi\rho}$.
The maximum initial field strength in the box corresponds to a minimum plasma beta parameter of $\beta = 10$, i.e. thermal pressure dominates over magnetic pressure.

The initial poloidal field is implemented as a uniform field in the vertical direction
\begin{eqnarray} \label{eq:poloidal_field}
	\textrm{B}_{\it{z}} = \textrm{B}_{0,\textit{z}}
	\label{eq:bz0}
\end{eqnarray}
for this field configuration the vertical Alfv\'en frequency is also constant everywhere, $\omega_{\textit{A},\textit{z}} = \textrm{v}_{\textit{A,z}}(2\pi/\textit{L}_{\textit{z}})$,  where $\textrm{v}_{\it{A,z}}=\textrm{B}_{\it{z}}/\sqrt{4\pi\rho}$ is the vertical Alfv\'en velocity and $\it{L}_{\it{z}}$ is the vertical extent of the box.

\begin{figure} 
 \centering
  \includegraphics[scale=0.3]{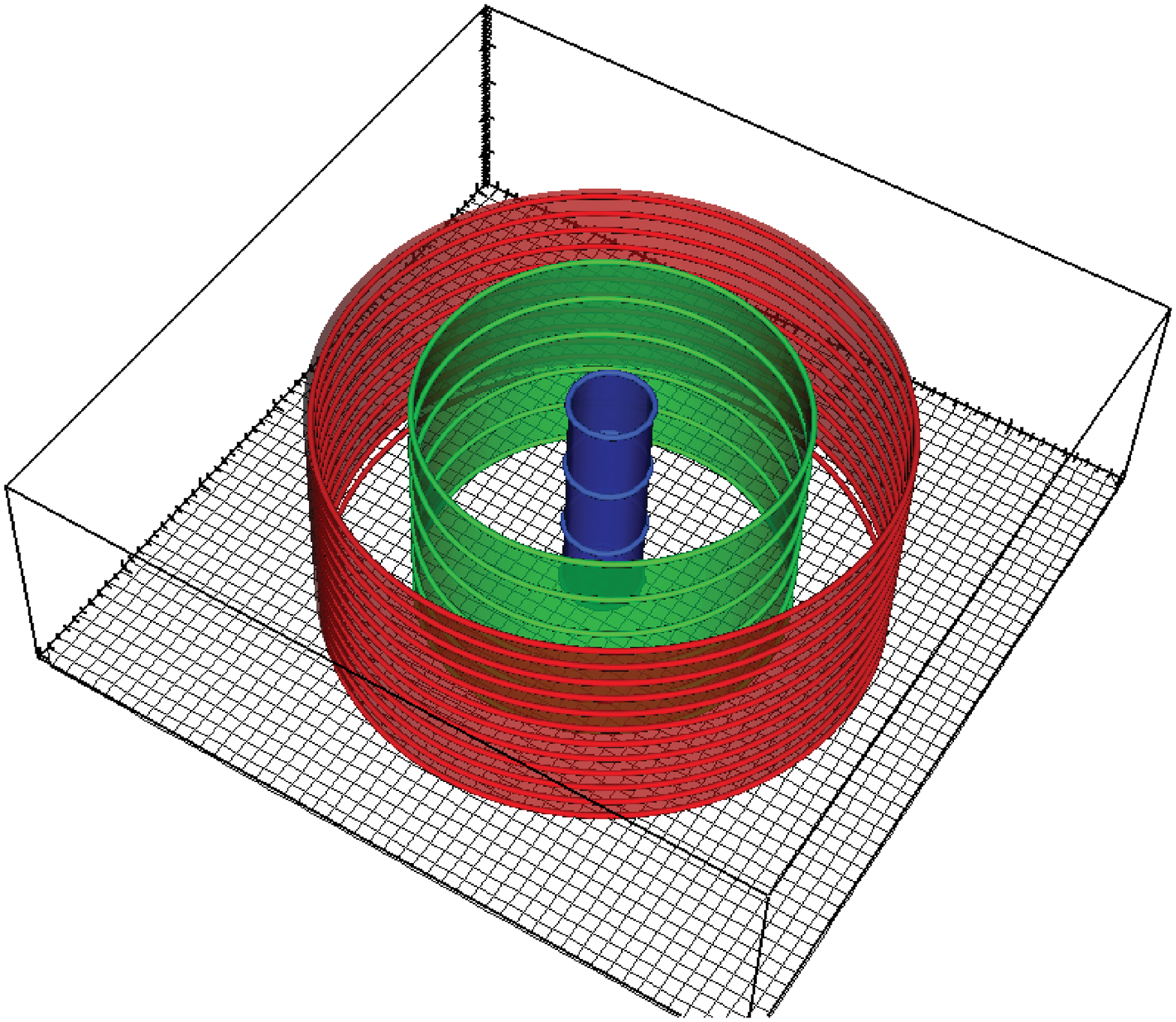}
  \includegraphics[scale=0.25]{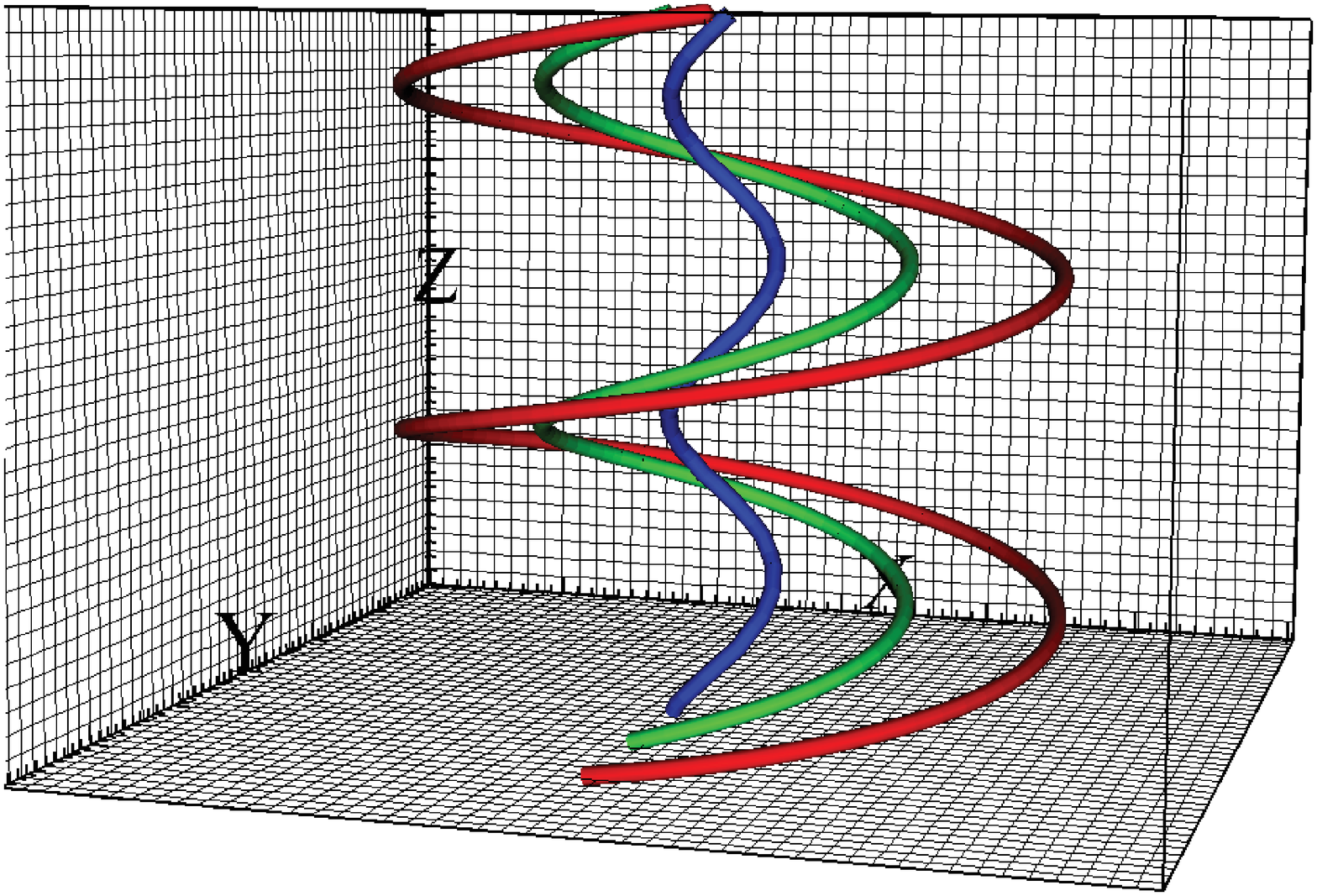}
  \caption{a) Initial purely toroidal magnetic configuration. The region of interest is between $0 \le \varpi \lesssim \varpi_{0}$ given that the field strength falls off at greater $\varpi$. 
   Colored areas represent surfaces of  equal distance to the axis. Field lines are also plotted; their density is related to the increase in the field strength proportional to the distance from the axis.
b) Mixed poloidal-toroidal field configuration. The field lines show a spiral morphology, for this particular case (with $p=1$) two spirals turns are completed in the box, i.e. toroidal Alfv\'en frequency is double the vertical Alfv\'en frequency $\omega_{\it{A}, \phi} = 2\omega_{\it{A},z}$.
}
 \label{fig:schematic_view}
\end{figure}

A visual example of the field configuration, for purely toroidal and mixed poloidal-toroidal fields, is shown in figure \ref{fig:schematic_view}. 

The initial conditions for the thermodynamic variables are determined by the equilibrium solutions found by setting $\partial/\partial t = \bf{v} = \eta = 0$ in equations (\ref{eq:momentum_eq})  to (\ref{eq:energy}), with the initial magnetic field
 given by equations (\ref{eq:magnetic_field}, \ref{eq:bz0} and \ref{eq:bphi}).
No height dependence is considered, therefore pressure, density and temperature depend only on the distance from the axis.
For a full derivation of these stability configurations see \citealt{Braithwaite06b} (section 3.3).

Four different cases for two different field configurations are simulated:
\begin{itemize}
	\item[1] rotating diffusive star,
	\item[2] perpendicular rotator,
	\item[3] rotating star with a quadratic radial field gradient: $p=2$, and
	\item[4] mixed poloidal-toroidal field.
\end{itemize}

\noindent\textbf{Case 1}: Rotating diffusive star with linear radial field gradient, $p=1$.
From (\ref{eq:toroidal_field}) we see that the toroidal Alfv\'en frequency varies as $\omega_{\rm A}\propto \varpi^{p-1}$ at low $\varpi$, so that it is approximately independent of location in the inner part of the computational box.
The magnetic and rotation axes are aligned.
We explore the parameter space of rotation-only, magnetic diffusion-only and the simultaneous action of rotation and diffusion. \\

\noindent\textbf{Case 2}: Perpendicular rotator with linear radial field gradient, $p=1$.
Rotation and magnetic axes are perpendicular.
We explore the rotation rate parameter space. \\

\noindent\textbf{Case 3:} Rotating star with a quadratic radial field gradient, $p=2$. 
The Alfv\'en frequency increases linearly with the distance to the axis.
For this case rotation and magnetic axes are aligned. 
We explore the rotation rate parameter space.\\

\noindent\textbf{Case 4:} Mixed poloidal-toroidal configuration.
The poloidal field is constant in the box, therefore the vertical Alfv\'en frequency is also constant.
The toroidal field has a linear radial field gradient, $p=1$, thus the azimuthal Alfv\'en frequency is also constant in the inner part of the box.
We explore the poloidal-to-toroidal field strength ratio parameter space. \\

Given that the initial configurations correspond to equilibrium solutions, an initial perturbation is included to trigger instability.

\subsection{Initial perturbation} \label{subsec:initial_perturbation}

The initial conditions in section \ref{subsec:initial_conditions} correspond to a perfect hydrostatic equilibrium, in order to see any instability grow we seed tiny perturbations in the velocity field.
Given the cylindrical symmetry of this problem, we can perturb the different azimuthal, $m$, and vertical, $n$, modes independently.
The shape and description of the perturbation used here are the same as used by \citet[][sec.~3.5]{Braithwaite06b}.
The perturbation has its maximum at $\varpi=0$ and decreases with increasing $\varpi$, almost vanishing at $\varpi=\varpi_{0}$.
There is no perturbation to the vertical velocity component. 

The initial perturbation has the form
\begin{align}
 \textbf{v} &= \sum_{m=0,n} V^{n}_{0} \frac{\varpi}{\varpi_{0}}\text{exp} \left (- \frac{5}{2} \frac{\varpi^2}{\varpi^2_{0} }\right) \text{cos}(k z) \textbf{e}_{\varpi} + \nonumber \\
 & \sum_{m \not=0,n}  V^{n}_{m} \text{exp}\left(-3\frac{\varpi}{\varpi_0}- \frac{\varpi^2}{\varpi^2_0}\right) [\text{cos}(k z-m\phi) \textbf{e}_{\varpi} + \text{sin}(k z-m\phi)\textbf{e}_{\phi}]. \nonumber
\label{eq:displacement_mmode}
\end{align}

The amplitude coefficients of the perturbed velocity field are normalized to the maximum Alfv\'en velocity in the box: $\textrm{V}_{\textit{m}}^{\textit{n}}/v_{A,\text{max}}=1\times10^{-8}$.
Such a value is large enough to trigger the Tayler instability and small enough to let us follow the linear growth phase for several growth timescales.

\subsection{Boundary conditions} \label{subsec:initial_perturbations}

The box represents a small volume of a radiative zone in a stellar interior.
Figure \ref{fig:Simulation_box} is a cartoon of the box location in star with respect to the magnetic field axis. 

The difference in length scales between the instability and the radius of the star are huge.
The instability is many orders of magnitude smaller\footnote{Up to 8 or 9 orders of magnitude assuming a $1 \rm{G}$ toroidal field in the Sun} than the radius of the star $(\it{l}_{\rm{r}} \ll \rm{R}_{*})$, these instabilities are local in the radial direction and global in the azimuthal direction, therefore different shells of the star are in effect disconnected from each other.

Periodic boundaries are the best option for our simulation, given that we want to avoid the selection of special locations in the stellar interior and to prevent additional effects such as current sheets or shear. 

\section{Results} \label{sec:results}

First, test cases without rotation or magnetic diffusion were simulated with field gradient $p=1$. The Tayler instability is properly traced for a number of growth timescales.  More physical ingredients are built up each step in order to approach our simple model with the conditions found in a radiative stellar interior.

For the analysis of the simulations we need a formalism for measuring the amplitude of the various azimuthal $m$ and vertical $n$ modes, keeping in mind that $n$ is the dimensionless vertical wavenumber; the $n=1$ mode corresponds to a wavelength equal to the height of the computational box. We extract information from two fields: the change in the magnetic field $\delta\textbf{B}=\textbf{B}-\textbf{B}(t=0)$, and the velocity field \textbf{v}. If we imagine the field to be frozen into the fluid, then $\delta \textbf{B}$ reflects the displacement ${\bm\xi}$, the integral of the velocity \textbf{v}.

We can now decompose the $\phi$ component of these fields using Fourier series in order to extract the amplitudes of the different azimuthal $m$ modes. For instance with the velocity field
\begin{eqnarray}
\textbf{v}(\varpi,\phi, \textit{z},\, t) = \frac{1}{2} \textbf{\textrm{A}}_{0} (\varpi, \textit{z},t) +\sum_{ \textit{m}=1}^{\textit{m}=\infty} \, \Re(\textbf{A}_{m}(\varpi,\textit{z},\textrm{t}) \, \textit{e}^{\textit{im}\phi}),
 \label{eq:first_fft}
\end{eqnarray}
where $\textbf{A}_{0}$ and $\textbf{A}_{m}$ are the Fourier coefficients given by
\begin{eqnarray*}
	\textbf{A}_{0}  (\varpi, \textit{z},t)  =   \frac{1}{\pi} \int_{-\pi}^{\pi} \textbf{v}(\varpi, \phi, z, t) \, {\rm d}\phi 	
\end{eqnarray*}
and
\begin{eqnarray*}
	\textbf{A}_{m} (\varpi, \textit{z},t)  =  \frac{1}{\pi} \int_{-\pi}^{\pi} \textbf{v} (\varpi, \phi, z, t) \, \cos(m\phi) \, {\rm d}\phi.
\end{eqnarray*} 

We then compute the amplitude of each $\textit{m}$ mode as

\begin{equation}
	\begin{split}
	\textrm{v}\textit{\_Amp}_{\textit{m}}(\rm{t}) =  \left( \frac{\int_{0}^{\varpi_{\textrm{max}}} \int_0^{L_z} \textbf{A}_{m}^{*} \, \textbf{A}_{m} \, 2\pi\varpi \, {\rm d}\varpi {\rm d}\rm{z} }{\varpi_{\rm max}^2 L_z \pi}\right)^{1/2},
	\label{eq:amplitude_mode}
	\end{split}
\end{equation}
where $\varpi_{\textrm{max}}$ is set to $\varpi_0/2$, as the magnetic field strength in the region between $\varpi=0 $ to $ \varpi_{0}/2$ is approximately proportional to the distance from the axis to the power $\rm{p}$, i.e. $B_\phi \propto \varpi^{\textrm{p}}$.
A similar procedure can be applied to the magnetic fields to derive the magnetic displacement amplitude, $\delta B\textrm{\_Amp}_m(t)$.

\begin{figure}
 \centering
 \includegraphics[scale=0.53]{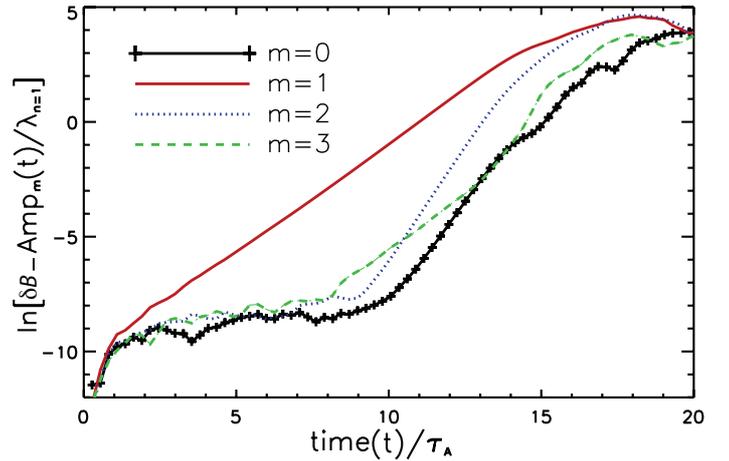}
 \caption{Amplitude of the first four $m$ modes as a function of time, extracted from the magnetic field $\delta B(t)$, and normalized to the wavelength corresponding to the $n=1$ vertical mode. The azimuthal $m=1$ mode, solid red line, is the only unstable mode. }
 \label{fig:amplitude_default_case}
\end{figure}

Figure \ref{fig:amplitude_default_case} shows the evolution of the amplitudes of the first three $m$ modes as a function of time.  The $m=1$ mode is the only unstable mode, as predicted above in (\ref{eq:stability_non-rotating}).  When the amplitude of the unstable $m=1$ mode is comparable to the wavelength, $\delta B\_Amp_m \approx \lambda_{n=1}$, the system goes into the non-linear regime. At this point energy can be transferred between modes and $ m\neq 1$ modes begin to grow. This is observed to happen around $t/\tau_{\rm A}\approx 10$ (where $\tau_{\rm A}=1/\omega_{\rm A}$) in figure \ref{fig:amplitude_default_case}.

The displacement amplitude computed using equation (\ref{eq:amplitude_mode}) contains the contribution of all vertical wavelengths. For further analysis, it will be necessary to measure the amplitudes of each individual vertical $n$ modes separately. A second Fourier decomposition, similar to the one applied in equation (\ref{eq:first_fft}), is performed in the vertical direction,
\begin{eqnarray}
\delta\textbf{v}(\varpi, \phi, z, t) = \frac{1}{2} \textbf{C}_{m, 0} \, (\varpi, t) + \sum_{n=1}^{n=\infty}\Re(\textbf{C}_{m, n}(\varpi,t) \, \textit{e}^{inz}),
\label{eq:second_fft}
\end{eqnarray}
 where $\textbf{C}_{m,0}$ and $\textbf{C}_{m, n}$ are the new Fourier coefficients given by:
\begin{equation*}
	\textbf{C}_{m,0} = \frac{1}{\pi L_{z}} \int_{0}^{L_{z}} \int_{-\pi}^{\pi} \delta \textbf{v} \, \cos(m\phi)   \, d\phi \, dz
\end{equation*} 
and
\begin{equation*}
	\textbf{C}_{m, n} = \frac{1}{\pi L_{z}} \int_{0}^{L_{z}} \int_{-\pi}^{\pi} \delta \textbf{v} \, \cos(m \phi) \, \cos(nz) \, {\rm d}\phi  \, {\rm d}z    
\end{equation*} 
and are used to compute the displacement amplitude for each individual pair of azimuthal $m$ and vertical $n$ modes, $\textrm{v}\textrm{\_Amp}_{\textit{m,n}}(t)$, similar to the procedure applied in equation  \ref{eq:amplitude_mode}.

\subsection{Case 1 -- rotating diffusive star} \label{subsec:case1}

We first simulated the effects of just rotation and just diffusion on their own, for the constant-Alfv\'{e}n-frequency case, $p=1$. Rotation is parallel to the magnetic axis. We compare our results to those presented by \citet{Braithwaite06b} and show that we obtained similar results. 

The rotation-only case has a threshold for the rotation rate equal to the Alfv\'en frequency, $\Omega_{\parallel}=\omega_{{\rm A},\phi}$. Rotation rates faster than that completely suppress the Tayler instability, presumably imposing some epicyclic motion of the perturbed fluid elements around an equilibrium point.
\begin{figure}
 \centering
 \includegraphics[scale=0.52]{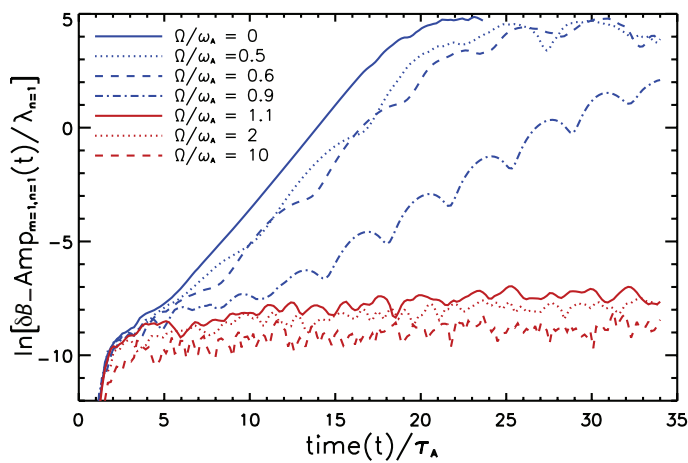}
 \caption{Amplitude $\delta B(t)$ of the $m=1$, $n=1$ mode as a function of time for various rotation rates. Blue lines correspond to rotation rates below the Alfv\'en frequency, and red lines correspond to rotation rates above the Alfv\'en frequency. }
 \label{fig:amp_vs_omega_p1}
\end{figure}
Figure \ref{fig:amp_vs_omega_p1} shows the amplitude of the $m=1$, $n=1$ mode in simulations with different $\Omega_{\parallel}$, ranging from $\Omega_{\parallel}/\omega_{\textrm{A},\phi} = 0 \, - \, 10$. It is clearly observed how the system is unstable for $\Omega_{\parallel} / \omega_{\textrm{A},\phi} < 1$ (blue lines), and is stabilized for rotation rates $\Omega_{\parallel} / \omega_{\textrm{A},\phi}>1$ (red lines). \\
In contrast, the diffusion-only case shows no threshold beyond which the Tayler instability is completely stabilized.

\begin{figure}
 \centering
  \includegraphics[scale=0.8]{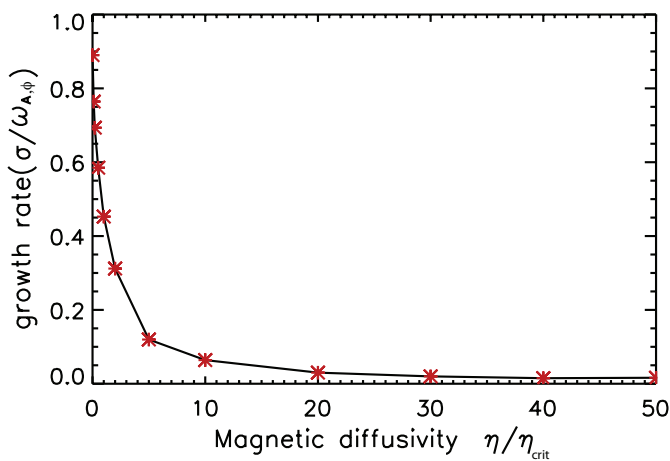}
 \caption{Growth rate of the Tayler instability for the $m=1$ and $n=1$ mode as a function of the relative magnetic diffusivity $\eta/\eta_{\rm crit}$ measured from the velocity field, $v\_{Amp}_{m, n}(t)$.}
 \label{fig:sigma_vs_eta}
\end{figure}
 Figure \ref{fig:sigma_vs_eta} shows the measured unstable growth rates in terms of $n_{\eta}$.
 It is observed that even at diffusivities as high as $n_{\eta}=50$; unstable growth is still present.

\begin{figure}
	\centering
  	\includegraphics[scale=0.6]{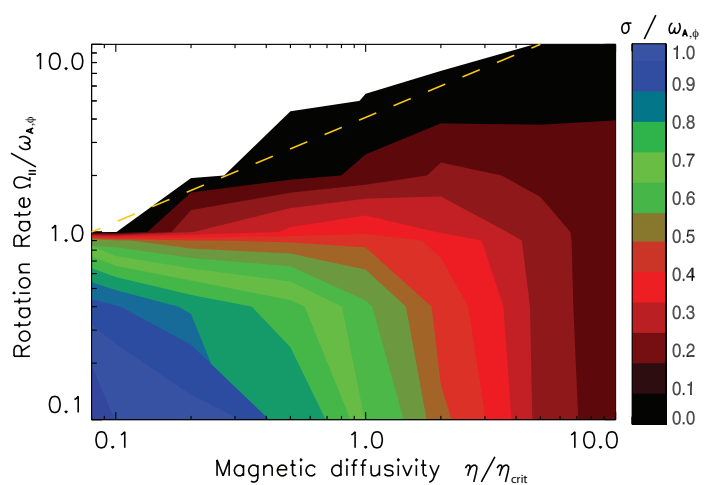}
 	\caption{Contours of mean growth rate plotted as a function of normalized rotation rate $n_{\Omega_{\parallel}} = \Omega_{\parallel}/\omega_{{\rm A},\phi}$ and the relative magnetic diffusion $n_{\eta} = \eta/\eta_{\textrm{crit}}$. 
 The white region represents stability, for at least $100 \tau_{A}$.
The color contours correspond to increasing growth rates of the instability, where black corresponds to small and blue to fast growth rates.
The yellow dashed line corresponds to the relation $n_{\Omega_{\parallel}} = 2n_{\eta} + 1$ separating the stable and unstable regions.}
 \label{fig:omega_eta}
\end{figure}

For the simultaneous action of both rotation and magnetic diffusion a total of 56 simulations were conducted.
Figure \ref{fig:omega_eta} summarizes the measured growth rates normalized to the Alfv\'en frequency, $\sigma/\omega_{\rm{A},\phi}$.

We consider a system to be stable if no unstable growth is observed for a hundred Alfv\'en crossing timescales, $100 \tau_A$, corresponding to the white region in figure \ref{fig:omega_eta}. We see that at very low magnetic diffusivities, left side of the figure, the system behaves similarly to the non-diffusive case, the system is stable for rotation rates $\Omega_{\parallel} > \omega_{A,\phi} $.
Now if we move from the stable region in a straight horizontal line towards the right side of the figure we go towards higher magnetic diffusive environments and penetrate a region of unstable growth. This is a very interesting result as one would {\it prima facie} expect that two stabilizing processes, such as rotation and magnetic diffusion, to collaborate in suppressing the instability, but instead the presence of magnetic diffusion disturbs the stabilization introduced by rotation.

Presumably this happens because the Coriolis force pushes a fluid element onto an epicyclic equilibrium motion; magnetic diffusion constantly removes energy from the magnetic field such that there is no longer sufficient energy to return the fluid element to an equilibrium epicycle, and it will slowly spiral outwards.

The dashed yellow line in figure \ref{fig:omega_eta} is a relation between $n_{\Omega_{\parallel}}$ and $n_{\eta}$. If we assume that the boundary dividing the stable and unstable regimes can be separated by a linear relation we obtain
\begin{equation}
n_{\Omega_{\parallel}} \approx 2n_{\eta} + 1.
\label{eq:stabilityrelation}
\end{equation}
It is not clear why the threshold should be here. The factor of 2 may come from the 2 in the Coriolis force itself.

Alternatively the stability condition can be expressed as
\begin{equation}
	\omega_{A,\phi} < \Omega_{\parallel} - \eta k^2.
	\label{eq:stabilityconditions}
\end{equation}
This is the stability condition for fast rotating, diffusive systems as a function of the perturbed wavenumber $k$.
The growth rate of the instability is in general a function of the Alfv\'en frequency, rotation rate, and magnetic diffusion rate, $\sigma=\sigma(\omega_{\rm{A}},\Omega_{\parallel},\eta\it{k}^2)$ for the non-stratified case.

\subsection{Case 2 -- perpendicular rotator} \label{subsec:case2}

\begin{figure}
	\centering
  \includegraphics[scale=0.65]{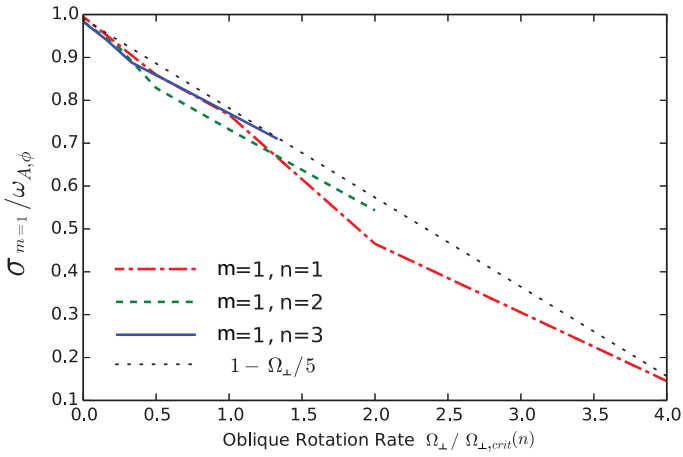}
 \caption{Growth rate of the Tayler instability, normalized to the Alfv\'en frequency, as a function of the perpendicular rotation rate, normalized to the predicted critical rotation rate for each $n$ mode.
The lines approximately on top of each other show the dependence of the damping on wavenumber is as predicted.}
 \label{fig:oblique_mn}
\end{figure}

We simulated a rotator perpendicular to the magnetic field axis of symmetry. 
The magnetic field is oriented in the $z$-axis, $\hat{\bf B} = \hat{e}_{z}$, and the rotation axis is oriented along the $\varpi(\phi=0)$-direction ${\bm\Omega}=\Omega_\perp(\phi=0){\bf e}_{\varpi}$. As in the previous section, the field gradient is $p=1$.  We expect rotation to have a different effect on different vertical wavelengths, as given by equation \ref{eq:critical_rotation_oblique}. Figure \ref{fig:oblique_mn} shows the growth rates of the first three vertical $n$ modes (all $m=1$) as a function of the rotation rate normalized to the critical rotation for each $n$ mode. We can observe how the growth rate of the instability is damped with increasing rotation rate and furthermore how this damping affects different vertical modes. It is however important to remember that the horizontal axis of the plot is normalized to the critical rotation rate for each vertical mode. This means that given a rotation rate of e.g. $\Omega_{\perp} = 4 \Omega_{\perp,{\rm crit}}(n=1) = 20 \omega_{{\rm A},\phi}$, the $n=1$ mode is damped by $\sim 90\%$, $n=2$ mode by $\sim 45\%$, and $n=3$ mode by $\sim 25\%$.
In agreement with our prediction that perpendicular rotation damps different vertical modes with a different strength.

Even if we did not see complete stability in our simulation, figure \ref{fig:oblique_mn} suggests that there is stabilization for each vertical mode at a rotation rates of $\Omega_{\perp} \approx 5 \Omega_{\perp,{\rm crit}}(n)$. 
Apparently our stability condition (\ref{eq:critical_rotation_oblique}) is missing a factor of order unity. This is not surprising, given that the damping effect should depend on the precise arrangement of the displacements within the mode.

\subsection{Case 3 -- quadratic radial field gradient} \label{subsec:case3}

We saw above that in the case where $p=1$, rotation can completely stabilize the magnetic field, at least where there is no diffusion. We now investigate the case with a radial field gradient of $p=2$, which is predicted to behave differently. 
The Alfv\'en frequency grows proportional to distance from the axis $\omega_{{\rm A},\phi}(\varpi)\propto\varpi$.
For this reason we expect that the instability grows faster at larger radii than close to the axis.
We introduce a new quantity, $\delta B\textrm{\_Amp}_{m, n}(\varpi,t)$, corresponding to the displacement amplitude, for a given pair of azimuthal and vertical modes, as a function of time and distance to the axis $\varpi$:
\begin{eqnarray}
\delta B\textit{\_Amp}_{m, n}(\varpi,\textrm{t})=\left( \frac{{\bf C}_{m,n}^{*}(\varpi, t){\bf C}_{m,n}(\varpi, t) }{\varpi^2 L_z \pi}\right)^{1/2}.
\label{eq:amplitude_mode_wavenumber_varpi}
\end{eqnarray}
This new quantity allows us to study the growth rate of the instability at different distances from the axis.

\begin{figure}[!h]
 \centering
\includegraphics[scale=0.55]{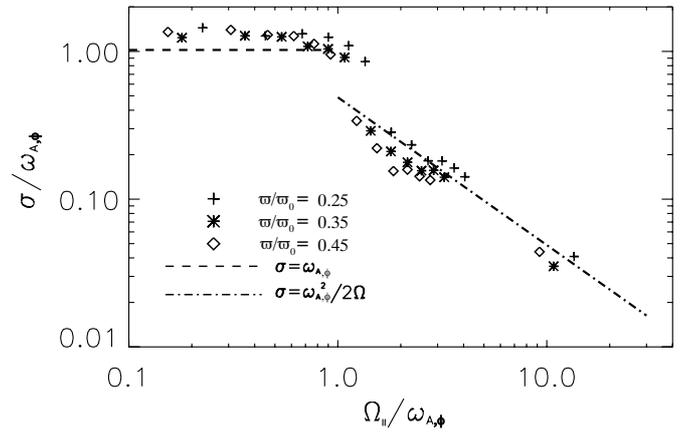}
 \caption{Growth rate $\sigma$ of the Tayler instability as a function of the solid-body rotation rate $\Omega_{\parallel}$, in the $p=2$ case.
Normalized local growth rates are given for three different distances from the axis, $\sigma(\varpi)$.
The dashed line corresponds to the predicted growth rate of the Tayler instability for slow rotators, $\sigma = \omega_{A,\phi}$.
The dotted-dashed line is the predicted behavior of the instability for fast rotators, $\sigma = \omega_{A,\phi}^2/2\Omega_{\parallel}$.}
 \label{fig:p2_fit_sigma_omega}
\end{figure}

Local growth rates, $\sigma(\varpi)$, are computed from the local amplitude displacements, $\delta B{\rm \_Amp}_{m, n}(\varpi,t)$. 
This quantity is then divided by the local Alfv\'en frequency, $\omega_{{\rm A},\phi}(\varpi)$, and the normalized local growth rate is obtained, $\sigma(\varpi)/\omega_{{\rm A},\phi}(\varpi)$.

Figure \ref{fig:p2_fit_sigma_omega} shows the behavior of these normalized local growth rates as a function of rotation rate.
The first main feature of the figure is that even for the very fast rotators, instability is still present.
This is in agreement with the prediction that instability is always present in fast rotating stars containing a purely toroidal field with a radial field gradient $p \gtrsim 3/2$, obtained from the stability conditions (\ref{eq:stability_rotating}).

There are two behaviors observed in this plot separated by a critical rotation rate at $\Omega_{\parallel}\approx\omega_{\rm{A},\phi}$.  In the slow rotation regime where $\Omega_{\parallel}/\omega_{{\rm A},\phi}\lesssim1$, the growth rate $\sigma\approx\omega_{\rm{A},\phi}$, whereas in the fast rotation regime where $\Omega_{\parallel}/\omega_{\rm{A},\phi}\gtrsim1$, we have instead $\sigma\approx\omega_{\rm{A},\phi}^2/2\Omega_{\parallel}$. Which is exactly what \citet{Pitts&Tayler85} and \citet{Spruit99} predicted a couple of decades ago.\footnote{Except that they neglected the factor of 2. At a latitude of $30^{o}$ this factor will go away anyway since the component of $\bm{\Omega}$ perpendicular to the spherical surfaces in which the fluid is moving around is half as much as at the poles.}

\subsection{Case 4 -- mixed poloidal-toroidal field} \label{subsec:case4}

This section is intended to shed some light in the stability of mixed poloidal-toroidal magnetic field configurations in radiative stellar interiors. We focus our analysis on the behavior of the instability known to be present in purely toroidal fields, but now in mixed poloidal-toroidal field configurations. As discussed in section \ref{subsec:mixed_field_effects} a poloidal field could damp, and if strong enough probably be able to suppress, the instability of a toroidal field. Tayler's (1973) stability conditions however are not valid for mixed field configurations.

We note that the results presented in this section should be reviewed carefully because of the appearance of an unknown, spurious mode in the simulations growing always at the smallest scale in the vertical direction\footnote{Despite the many vertical resolutions tested: 8, 16, 32, 64 and 96 these perturbations, with vertical wavelength corresponding to that of the Nyquist frequency, was present in the $m = 1$ mode with a growth rate $\sigma_{\rm{unk}} \approx \omega_{\rm A}/2$.}. Such a mode was absent in purely poloidal and purely toroidal field simulations. We work under the hypothesis that any well resolved unstable mode growing faster than this unknown perturbation is real, and relevant for our analysis. The nature of the unknown mode will be studied in detail in a forthcoming paper and it must be established whether it is astrophysically relevant. \\

We first introduce some important definitions to avoid misunderstandings; the azimuthal Alfv\'en frequency, $\omega_{{\rm A},\phi} = \textrm{v}_{{\rm A},\phi}/\varpi$,  is calculated with the toroidal Alfv\'en velocity. In the initial conditions, the radial field gradient of the toroidal field is chosen for simplicity to grow linearly with the distance from the axis, $p=1$, therefore the Alfv\'en frequency is constant throughout the volume of interest. The vertical Alfv\'en frequency $\omega_{\rm{A},z}$ is calculated with the vertical Alfv\'en velocity, $\textrm{v}_{\rm{A},z}$ and is given by
\begin{equation}
	\omega_{{\rm A},z} = \frac{\textrm{v}_{{\rm A},z}}{L_z} = \frac{B_z}{L_z\sqrt{4\pi\rho}}.
\end{equation}
The poloidal field strength is constant everywhere, $B_z = B_0$, thus the vertical Alfv\'en velocity and the vertical Alfv\'en frequency are constant. \\

For this field configuration field lines follow spirals as shown in figure \ref{fig:schematic_view}b. The number of spirals a field line completes in the box depends on the relative strengths of the toroidal and poloidal fields. The more complete spirals a field line draws, the stronger the toroidal field is. 
We call this number of spirals in the box the pitch--number, $P_n$, which is given by
\begin{equation}
	P_n = \frac{\omega_{{\rm A},\phi}}{2\pi\omega_{{\rm A},z}};
\end{equation}
re-writing the instability condition (\ref{eq:poloidal_stability}) in terms of the pitch--number and solving for the $\textit{n}=1$ mode, we obtain that instability is expected to be present for $P_n > 1$. We note that $P_n$ is only independent of $\varpi$ in the special case $p=1$, which is what we use here. 

\begin{figure}
\centering
  \includegraphics[scale=0.58]{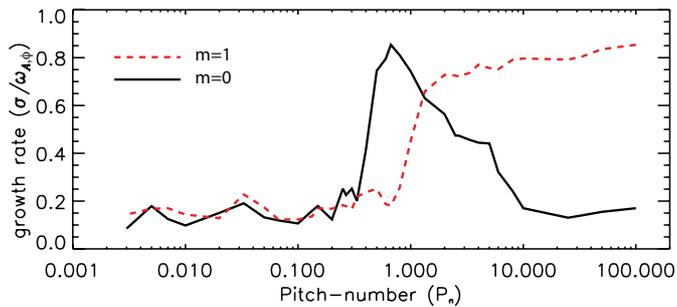}
 \caption{Growth rate of the first two azimuthal $\textit{m}$ modes as a function of the pitch--number. The solid black line corresponds to the $\textit{m}=0$ mode, stable at very high and low pitch--numbers but with a peak in unstable growth near the critical pitch--number, $P_n \approx 1$. The red dashed line corresponds to the $\textit{m}=1$ mode, unstable for very high pitch--numbers and stabilized near the critical pitch--number, $P_{n} \approx 1$.}
 \label{fig:Bz_sigma}
\end{figure}

Figure \ref{fig:Bz_sigma} shows the behavior of the first two azimuthal modes, $m=0$ and $m=1$, as a function of pitch--number. The $m=1$ mode has a growth rate on the order of the Alfv\'en frequency for a predominantly toroidal field configuration, i.e. large pitch--number, and drops suddenly when the pitch--number reaches a critical value. This drop of the growth rate agrees with the stability conditions (\ref{eq:poloidal_stability}). We note that no complete suppression of the instability is observed, even for very small pitch--numbers. However, these small growth rates are the effect of non-linear interaction with the spurious modes.

The second interesting mode shown in figure \ref{fig:Bz_sigma} is the $m=0$ mode.  This mode was marginally stable for the purely toroidal field (\ref{eq:stability_non-rotating}) and it is being perturbed in the mixed configuration. Growth of the $m=0$ mode begins to appear with increasing poloidal field strength, reaching a maximum near the critical poloidal-to-toroidal field ratio, $P_n\approx1$.

The presence of this unstable mode, not predicted analytically, might be important for the hydromagnetic dynamo theory, given that this instability triggers axisymmetric vertical motions which it is often claimed are necessary for closure of the dynamo loop. We will investigate the importance of this mode in a future paper.

\section{Discussion and summary} \label{sec:conclusions}

Solid-body rotation, magnetic diffusion and field configuration are expected to have a strong influence on the presence and evolution of hydro-magnetic instabilities in toroidal fields in radiative stellar interiors. We present a set of 3D MHD simulations of purely toroidal and of mixed poloidal-toroidal magnetic fields in non-convective stellar interiors.  A combination of various radial field strength gradients ($p={\rm d}\ln B/{\rm d}\ln \varpi$), solid body rotation rates ($\Omega$), magnetic diffusivities ($\eta$), and poloidal-to-toroidal field strengths ($P_n$) is scoped.

We observe the presence of the kink type $m = 1$ Tayler instability in purely toroidal fields. We measure its growth rate, and compare our results with the various analytical results for non-rotating and non-diffusive \citep{Tayler73}, fast rotating \citep{Acheson78,Pitts&Tayler85}, highly diffusive \citep{Spruit99}, and rotating-diffusive \citep{Pitts&Tayler85,Spruit99} radiative stellar interiors. We also study the development of instabilities in a mixed poloidal-toroidal field configuration. \\

Magnetic diffusion is not able to suppress the occurrence of the instability even at high diffusivity, in accordance with the results of \citet{Braithwaite06a}. 

{\bf Diffusion and rotation parallel to magnetic axis:} If the magnetic and rotation axes are parallel, fast rotation ($\Omega_{\parallel}>\omega_{{\rm A},\phi}$) is able to stabilize the configuration (see figure \ref{fig:amp_vs_omega_p1} ), where $\omega_{{\rm A},\phi}$ is the azimuthal Alfv\'en frequency. This is in the case where the field gradient $p=1$.
However as soon as magnetic diffusion is included, the stability conditions now depend on a combination of the rotation rate, the magnetic field strength and the magnetic diffusion rate given by equation \ref{eq:stabilityconditions}, Which suggests that very fast rotation can stabilize the system only if the magnetic diffusion rate does not subtract energy fast enough such that Coriolis force is unable maintain the fluid elements in stable epicyclic motions.
{\bf Perpendicular rotation:} If the magnetic field axis and the rotation axis are perpendicular, the damping strength introduced by rotation is different for different vertical wavelengths. The damping effect is stronger for small vertical wavenumbers $k$ compared to large wavenumbers because of the larger deformation occurring perpendicular to the rotation axis at smaller wavenumbers. 
It is found that the instability is suppressed if $\Omega_{\perp}\gtrsim 5\omega_{{\rm A},\phi}\varpi_0k$, where $\varpi_0$ is the length scale of the motions in the horizontal direction. \\

{\bf Radial field gradient:} To analyze the effects of the radial field gradient and rotation in the instability we investigated two cases, the linear radial field gradient, $p=1$ and the quadratic radial field gradient, $p=2$.
As described above, in the $p=1$ case (without diffusion) stability is reached for fast rotation (meaning faster than the azimuthal Alfv\'en frequency).
In this $p=2$ case, very fast rotation is never able to stabilize the configuration.
Linear growth rates behave broadly as predicted by \citet{Pitts&Tayler85}, where for slow rotators the growth rate is roughly equal to the Alfv\'en frequency, whilst for fast rotators, the growth rate is damped by a factor of $\omega_{\rm{A},\phi}/2\Omega_{\parallel}$. It is probably safe to assume that this is also the case for all values of $p$ above $3/2$, as originally predicted. We have therefore provided a check on the theory, useful in light of the fact that the originators of the analytic results were, we think it is fair to say, less than 100\% certain they were correct.

{\bf Mixed poloidal-toroidal field:} Finally we simulated a mixed poloidal-toroidal field, and in particular the stabilizing influence of a uniform poloidal field. In the $p=1$ case, 
where in the absence of a poloidal field, only the $m=1$ mode is unstable, we find that the $m=1$ mode is indeed suppressed by the poloidal field, and that the stability criterion is where it is predicted to be. Interestingly, we also find that the $m=0$ mode is unstable in the vicinity of this threshold.
\\

The presence of the unstable axisymmetric $m=0$ mode in the mixed poloidal-toroidal field configuration may be quite important. In the case where the field gradient is such that it is marginally stable in the purely toroidal case, it becomes dominant in the vicinity of the threshold of stabilization by the poloidal field, which of course is precisely where the dynamo is likely to be. It opens the possibility for axisymmetric turbulent radial motions which affect the mixing and the regeneration of the poloidal field. It is often claimed by various members of the MHD community that a poloidal field cannot be renewed without the $m=0$ mode; whether this claim is correct is, if the $m=0$ mode is present and perhaps even dominant, no longer relevant.

Indeed, as far as the field gradient $p$ is concerned, in the fast-rotating (which is expected to occur much more frequently in nature than the slowly rotating)  adiabatic case we can summaries behavior in terms of the threshold: where $p<1$ all modes are stable; where $1<p<3/2$ just the $m=0$ mode is unstable; where $p>3/2$ the $m=1$ mode is also unstable, with thresholds at higher values of $p$ introducing instability at progressively higher values of $m$. However, these thresholds are probably of little practical importance. Whilst it is often said that the Tayler instability is dominated by the $m=1$ mode because it is that mode which is in some sense the first to become unstable, this is misleading: only in the rather special circumstance of $-1/2<p<1$ and $\Omega_{\parallel}<\omega_{\rm A}$ is the $m=1$ mode the only one present. Firstly, the second of those two conditions is rarely fulfilled in nature. Secondly, in the context of the fluctuating field envisaged in the Tayler-Spruit dynamo model, the field gradient $p$ will have values all the way from $p=-\infty$ to $p=+\infty$. It seems perfectly plausible that the dynamo could be dominated by the $m=0$ mode, or indeed by $m=20$. Roughly speaking, the instability is present where $p$ is positive and absent where it is negative: the dynamo can therefore be expected to be intermittent in some sense.

\begin{acknowledgements}
	The authors would like to thank Alfio Bonanno and Henk Spruit for fruitful discussions. The analysis and plots presented here were performed using VisIt (Visualization Tool) and Ipython notebook, and the bibliography was put together thanks to the Astrophysical Database ADS and Mendeley.
\end{acknowledgements}

\bibliographystyle{aa} %
\bibliography{new_Toroidal_field_stability}

\begin{thebibliography}{32}
\expandafter\ifx\csname natexlab\endcsname\relax\def\natexlab#1{#1}\fi

\bibitem[{{Acheson}(1978)}]{Acheson78}
{Acheson}, D.~J. 1978, Royal Society of London Philosophical Transactions
  Series A, 289, 459

\bibitem[{{Akg{\"u}n} {et~al.}(2013){Akg{\"u}n}, {Reisenegger}, {Mastrano}, \&
  {Marchant}}]{Akgun13}
{Akg{\"u}n}, T., {Reisenegger}, A., {Mastrano}, A., \& {Marchant}, P. 2013,
  \mnras, 433, 2445

\bibitem[{{Babcock}(1947)}]{Babcock47}
{Babcock}, H.~W. 1947, \apj, 105, 105

\bibitem[{{Bonanno} \& {Urpin}(2013{\natexlab{a}})}]{Bonanno&Urpin2013a}
{Bonanno}, A. \& {Urpin}, V. 2013{\natexlab{a}}, \apj, 766, 52

\bibitem[{{Bonanno} \& {Urpin}(2013{\natexlab{b}})}]{Bonanno&Urpin2013c}
{Bonanno}, A. \& {Urpin}, V. 2013{\natexlab{b}}, \mnras, 431, 3663

\bibitem[{{Bonanno} \& {Urpin}(2013{\natexlab{c}})}]{Bonanno&Urpin2013b}
{Bonanno}, A. \& {Urpin}, V. 2013{\natexlab{c}}, \aap, 552, A91

\bibitem[{{Braithwaite}(2006{\natexlab{a}})}]{Braithwaite06a}
{Braithwaite}, J. 2006{\natexlab{a}}, Astronomy and Astrophysics, 449, 451

\bibitem[{{Braithwaite}(2006{\natexlab{b}})}]{Braithwaite06b}
{Braithwaite}, J. 2006{\natexlab{b}}, Astronomy and Astrophysics, 453, 687

\bibitem[{{Braithwaite}(2007)}]{Braithwaite07}
{Braithwaite}, J. 2007, Astronomy and Astrophysics, 469, 275

\bibitem[{{Braithwaite}(2008)}]{Braithwaite08}
{Braithwaite}, J. 2008, Monthly Notices of the RAS, 386, 1947

\bibitem[{{Braithwaite}(2009)}]{Braithwaite09}
{Braithwaite}, J. 2009, Monthly Notices of the RAS, 397, 763

\bibitem[{{Braithwaite} \& {Nordlund}(2006)}]{Braithwaite&Nordlund05}
{Braithwaite}, J. \& {Nordlund}, {\AA}. 2006, Astronomy and Astrophysics, 450,
  1077

\bibitem[{{Braithwaite} \& {Spruit}(2004)}]{Braithwaite&Spruit04}
{Braithwaite}, J. \& {Spruit}, H.~C. 2004, Nature, 431, 819

\bibitem[{{Chaplin} {et~al.}(2001){Chaplin}, {Elsworth}, {Isaak}, {Marchenkov},
  {Miller}, \& {New}}]{Chaplin01}
{Chaplin}, W.~J., {Elsworth}, Y., {Isaak}, G.~R., {et~al.} 2001, \mnras, 327,
  1127

\bibitem[{{Collins}(1963)}]{Collins63}
{Collins}, II, G.~W. 1963, \apj, 138, 1134

\bibitem[{{Heger} {et~al.}(2003){Heger}, {Woosley}, \& {Langer}}]{Heger03}
{Heger}, A., {Woosley}, S.~E., \& {Langer}, N. 2003, in IAU Symposium, Vol.
  212, A Massive Star Odyssey: From Main Sequence to Supernova, ed. K.~{van der
  Hucht}, A.~{Herrero}, \& C.~{Esteban}, 357

\bibitem[{{Lander} \& {Jones}(2011)}]{Lander&Jones2011}
{Lander}, S.~K. \& {Jones}, D.~I. 2011, \mnras, 412, 1394

\bibitem[{{Maeder} \& {Meynet}(2003)}]{Maeder03}
{Maeder}, A. \& {Meynet}, G. 2003, AAP, 411, 543

\bibitem[{{Mestel}(1953)}]{Mestel53}
{Mestel}, L. 1953, MNRAS 113,716

\bibitem[{{Nordlund} \& {Gaalsgard}(1995)}]{stagger}
{Nordlund}, A. \& {Gaalsgard}, K. 1995

\bibitem[{{Pitts} \& {Tayler}(1985)}]{Pitts&Tayler85}
{Pitts}, E. \& {Tayler}, R.~J. 1985, Monthly Notices of the RAS, 216, 139

\bibitem[{{Ram{\'{\i}}rez-Agudelo} {et~al.}(2013){Ram{\'{\i}}rez-Agudelo},
  {Sim{\'o}n-D{\'{\i}}az}, {Sana}, {de Koter}, {Sab{\'{\i}}n-Sanjul{\'{\i}}an},
  {de Mink}, {Dufton}, {Gr{\"a}fener}, {Evans}, {Herrero}, {Langer}, {Lennon},
  {Ma{\'{\i}}z Apell{\'a}niz}, {Markova}, {Najarro}, {Puls}, {Taylor}, \&
  {Vink}}]{Ramirez-Agudelo}
{Ram{\'{\i}}rez-Agudelo}, O.~H., {Sim{\'o}n-D{\'{\i}}az}, S., {Sana}, H.,
  {et~al.} 2013, \aap, 560, A29

\bibitem[{{R{\"u}diger} {et~al.}(2014){R{\"u}diger}, {Schultz}, \&
  {Kitchatinov}}]{Rudiger2014}
{R{\"u}diger}, G., {Schultz}, M., \& {Kitchatinov}, L.~L. 2014, ArXiv e-prints

\bibitem[{{Spruit} \& {Phinney}(1998)}]{Spruit&Phinney98}
{Spruit}, H. \& {Phinney}, E.~S. 1998, Nature, 393, 139

\bibitem[{{Spruit}(1998)}]{Spruit98}
{Spruit}, H.~C. 1998, Astronomy and Astrophysics, 333, 603

\bibitem[{{Spruit}(1999)}]{Spruit99}
{Spruit}, H.~C. 1999, Astronomy and Astrophysics, 349, 189

\bibitem[{{Spruit}(2002)}]{Spruit02}
{Spruit}, H.~C. 2002, Astronomy and Astrophysics, 381, 923

\bibitem[{{Tayler}(1957)}]{Tayler57}
{Tayler}, R.~J. 1957, Proceedings of the Physical Society B, 70, 31

\bibitem[{{Tayler}(1973)}]{Tayler73}
{Tayler}, R.~J. 1973, Monthly Notices of the RAS, 161, 365

\bibitem[{{Williamson}(1980)}]{Williamson80}
{Williamson}, J.~H. 1980, Journal of Computational Physics, 35, 48

\bibitem[{{Woosley} {et~al.}(2002){Woosley}, {Heger}, \& {Weaver}}]{Woosley02}
{Woosley}, S.~E., {Heger}, A., \& {Weaver}, T.~A. 2002, Reviews of Modern
  Physics, 74, 1015

\bibitem[{{Zahn}(1992)}]{Zahn92}
{Zahn}, J.-P. 1992, Astronomy and Astrophysics, 265, 115

\end{thebibliography}

\end{document}